\documentclass[english,twocolumn]{revtex4-1}
\usepackage{amsmath}
\usepackage{mathptmx}
\usepackage{helvet}
\usepackage{newtxmath}
\usepackage[T1]{fontenc}
\usepackage[latin9]{inputenc}
\usepackage{graphicx}

\makeatletter

\usepackage{hyperref}

\newcommand{\expv}[1]{\left<#1\right>}
\newcommand{\ER}{Erd\H{o}s--R\'enyi}
\newcommand{\Var}[1]{\mathrm{Var}\left[#1\right]}

\makeatother

\usepackage{babel}
\begin{document}

\title{Generalization of the small-world effect on a model approaching the
\ER{} random graph}

\author{Benjamin F. Maier}
\email{bfmaier@physik.hu-berlin.de}

\affiliation{Robert Koch-Institute, Nordufer 20, D-13353 Berlin, Germany}

\affiliation{Department of Physics, Humboldt-University of Berlin, Newtonstraße
15, D-12489 Berlin, Germany}
\begin{abstract}
The famous Watts--Strogatz (WS) small-world network model does not
approach the \ER{} (ER) random graph model in the limit of total
randomization which can lead to confusion and complicates certain
analyses. In this paper we discuss a simple alternative which was
first introduced by Song and Wang, where instead of rewiring, edges
are drawn between pairs of nodes with a distance-based connection
probability. We show that this model is simpler to analyze, approaches
the true ER random graph model in the completely randomized limit,
and demonstrate that the WS model and the alternative model may yield
different quantitative results using the example of a random walk
temporal observable. An efficient sampling algorithm for the alternative
model is proposed. Analytic results regarding the degree distribution,
degree variance, number of two-stars per node, number of triangles
per node, clustering coefficient, and random walk mixing time are
presented. Subsequently, the small-world effect is illustrated by
showing that the clustering coefficient decreases much slower than
an upper bound on the message delivery time with increasing long-range
connection probability which generalizes the small-world effect from
informed searches to random search strategies. Due to its accessibility
for analytic evaluations, we propose that this modified model should
be used as an alternative reference model for studying the influence
of small-world topologies on dynamic systems as well as a simple model
to introduce numerous topics when teaching network science.
\end{abstract}
\maketitle

\section{Introduction}

When Watts and Strogatz published their 1998 paper ``Collective dynamics
of 'small-world' networks'' \citep{watts_collective_1998}, it had
a phenomenal influence on the field of complex systems and was one
of the defining studies for the following success of network science
to emerge as an interdisciplinary field. Not only was it the first
of a succession of studies \citep{kleinberg_small-world_2000,watts_identity_2002,newman_renormalization_1999,barrat_properties_2000}
trying to explain the small-world effect as based on Milgrim's ``six
degrees of separation'' experiment \citep{travers_experimental_1969},
it introduced a simple and intuitive network model which had, at its
core, the defining properties to obtain a ``complex'' system. Starting
with a regular, locally connected structure, a rewiring process introduces
long-range edges until it ends in a completely randomized state,
thus interpolating between two well-studied physical systems: a crystal
and disorder. For even small amounts of rewired contacts, the probability
that two neighbors are connected (typically large in social networks
\citep{newman_networks:_2010}) hardly changes, while almost immediately,
short paths between individual nodes appear, explaining how social
networks can be both: highly clustered but with a small amount of
necessary steps to reach one node from another.

This network model, based on rewiring edges, is widely used in the
network literature, often to explore the influence of the small-world
effect on the outcome of dynamic processes taking their course on
the network. One important feature of the model is that the mean degree,
i.e. the average number of nodes a node is connected to, is constant,
which is a first-order control parameter for a variety of dynamic
systems based on, e.g.~random walks or epidemic spreading \citep{barrat_dynamical_2008}.
Within this kind of research, people sometimes argue that one limit
of the rewiring process reproduces \emph{the} random graph, i.e.\ the
\ER{} $G(N,p)$ random graph model (as, for instance, in \citep{barabasi_mean-field_1999,wiedermann_mapping_2017,porter_small-world_2012}),
where $N$ is the number of nodes and $p$ is the probability that
any two nodes are connected. However, due to the model's definition,
the maximally randomized Watts-Strogatz model does \emph{not} actually
equal the \ER{} model. More often, references to the disorded limit
do not directly mention the \ER{} model as a limit but are ambiguous
in their wording and thus easily misinterpretable, see e.g.~\citep{jalan_random_2007,delgado_emergence_2002,bassett_small-world_2006,sporns_networks_2011,preciado_synchronization_2005,miklas_exploiting_2007,giacobini_takeover_2005}.
Likewise, the original rewiring procedure, where each node rewires
each of its edges to its \emph{rightmost} neighbors with probability
$p_{r}$ is similarly misinterpreted, to mean just rewiring any edge
\footnote{Until recently, even the model's Wikipedia page was clearly ambiguous
about its rewiring procedure.}. While the differences or slight variations of the model might not
be influential for some dynamics, in others they can cause clear deviations
from expected results in random graphs (see e.g.
Sec.~\ref{sec:model_definition}),
thus potentially leading to confusion or faulty interpretations.
The model is part of virtually every network science curriculum, however,
actually calculating the clustering coefficient, the degree distribution
or the small-world effect with pen and paper \footnote{or chalk and blackboard, depending on the preference}
is often omitted since these observables or effects are complicated
to evaluate. We argue that a model where those properties can be easily
evaluated without the aid of a computer and actually reproduce formerly
derived results from the \ER{} model might keep students more engaged
and trained to calculate properties of other network models.

A model which solves the problems discussed above has been introduced
by Song and Wang \citep{song_simple_2014}. Within their study, they
showed that sampling edges from a distance-based connection probability
eases the evaluations of e.g.~the degree distribution and the clustering
coefficient. In this paper, we reformulate and discuss this modified
model, propose an efficient sampling algorithm, extend the evaluation
of degree distribution and clustering coefficient to other network
properties, and show how it can be used to explain the small-world
effect analytically by comparing the clustering coefficient to an
upper bound of the message delivery time. Since the shortest path
length equals the delivery time of an optimal search process between
two nodes, the result presented here generalizes the small-world effect
to random search strategies.
\begin{figure}
\centering{}\includegraphics[width=1\columnwidth]{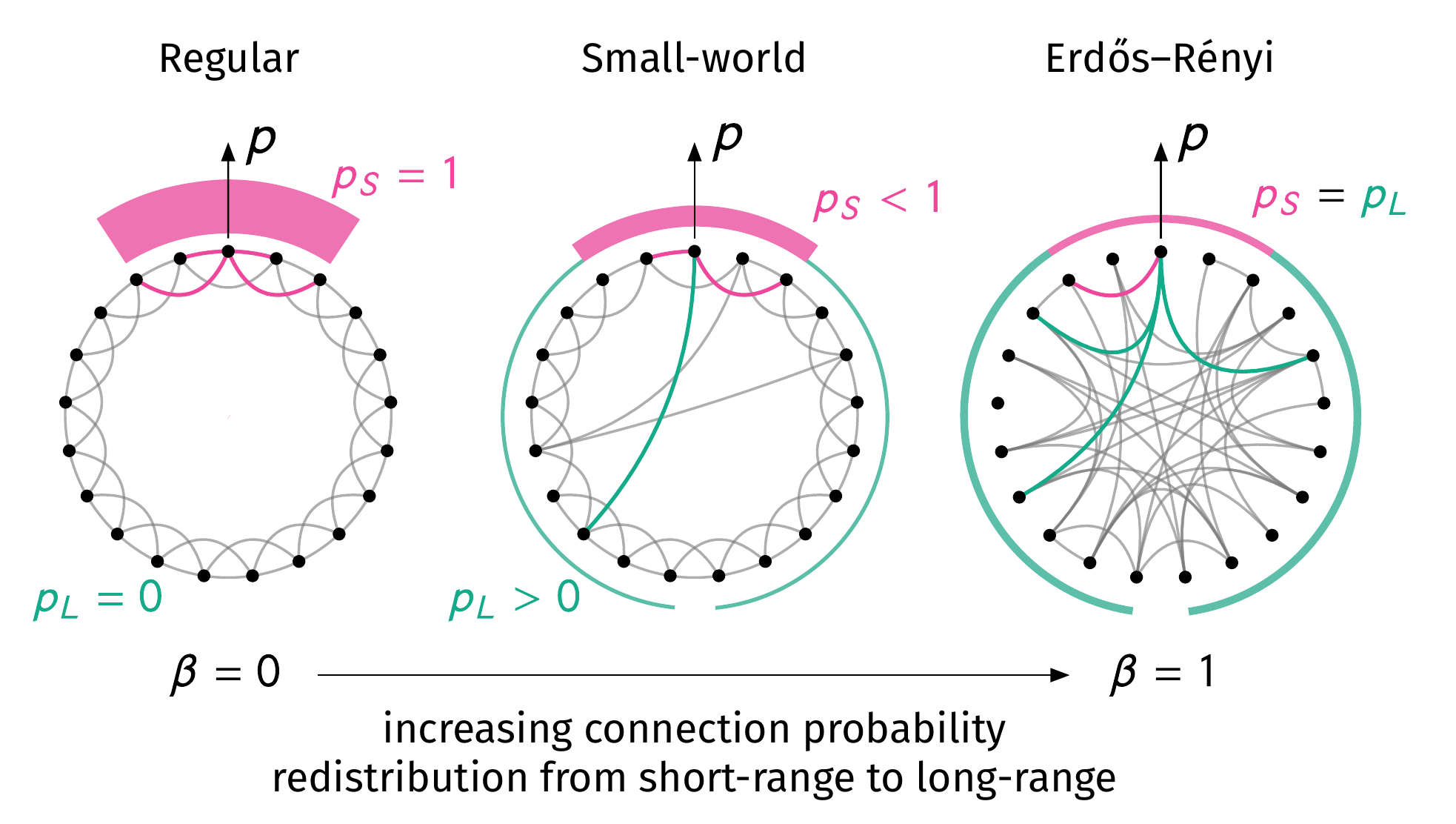}\caption{\label{fig:model_definition}Schematic representation of the alternative
small-world model as introduced in \citep{song_simple_2014} and discussed
in this paper. Much like in the original model, we start with $N$
nodes placed equidistantly on a ring. However, instead of rewiring,
each pair of nodes is connected with distance-based probability $p_{d}$
where $d$ is their minimal distance on the ring. Within distance
$d\protect\leq k/2$, nodes are connected with short-range probability
$p_{S}$. For larger distances, nodes are connected with long-range
probability $p_{L}=\beta p_{S}$. With increasing redistribution parameter
$0\protect\leq\beta\protect\leq1$ connection probability is redistributed
from the short-range regime to the long-range regime while the mean
degree $k$ is kept constant. Hence at $\beta=0$ the short-range
probability is unity while the long-range probability is zero which
produces a $k$-nearest neighbor lattice. With increasing $\beta$,
long-range ``short-cuts'' become more probable until at $\beta=1$
both connection probabilities are equal and thus the model becomes
equal to the \ER{} model.}
\end{figure}

\section{Results}

\subsection{Model definitions and differences}

\label{sec:model_definition}In the original model $N$ nodes are
positioned equidistantly on a ring and subsequently \emph{locally}
connected, i.e.\ connected to nodes in their vicinity with maximal
lattice distance $d\leq k/2$ where $k$ is an even positive integer.
In this state each node has degree $k$, where ``degree'' refers
to the number of neighbors of a node. For the rewiring process, each
node rewires its connections to its $k/2$ rightmost neighbors to
any other node in the network with probability $p_{r}$. It is easy
to see that at the randomized limit of $p_{r}=1$, each node has minimum degree $k/2$. Furthermore,
an original edge connected to a node $u$ has been rewired and can
only exist if it is reproduced by another rewiring event based on
its corresponding rightmost neighbor. This implies that at $p_{r}=1$
an original edge exists with probability $1/(N-1)$. Both these properties
lead to conceptual deviations from the \ER{} model in which each
edge exists with probability $p_{\mathrm{ER}}=k/(N-1)$ and nodes
may have degree $<k/2$.

In a variant of the modified model by Song and Wang \citep{song_simple_2014}
presented in the following, edges posses an inherent probability to
exist, which varies for \emph{short-range} ($S$) and\emph{ long-range}
($L$) contacts. A potential contact between nodes $(i,j)$ is considered
to be short-ranged if their distance in periodic boundary conditions
is $d(i,j)\leq k/2$; it exists with probability $p_{S}$. It is considered
long-range if $d(i,j)>k/2$ and exists with probability $p_{L}$.
The distance is computed as $d(i,j)=\min(|j-i|,N-|j-i|)$. In short,
two nodes with lattice distance $d$ are connected with probability
\[
p_{d}=\begin{cases}
p_{S,} & \mathrm{if\ }d\leq k/2,\\
p_{L}, & \mathrm{otherwise.}
\end{cases}
\]
Hence, if $p_{S}=1$ and $p_{L}=0$, the model produces a structure
which is equal to the original model's starting point, a one-dimensional
$k$-nearest neighbor lattice. On the other hand, if $p_{S}=p_{L}\equiv p$,
each edge exists with probability $p$ and hence the model reproduces
the $G(N,p)$ random graph. We can fix the mean degree by noticing
that it is composed of a short-range degree $\expv{k_{S}}$ and a
long-range degree $\expv{k_{L}}$. Each node has $k$ potential short-range
neighbors and $N-1-k$ potential long-range neighbors. Thus, its expected
degree is 
\begin{equation}
p_{S}k+p_{L}(N-1-k)=\expv{k_{S}}+\expv{k_{L}}=\expv{k}\equiv k.\label{eq:expected_degree}
\end{equation}
To keep the mean degree constant, we introduce a control parameter
$\beta$ which controls the trade-off of connection probability in
the short- and long-range regimes such that $p_{L}=\beta p_{S}.$
Note that at $\beta=0$, we have $p_{L}=0$ such that from Eq.~(\ref{eq:expected_degree})
it follows that $p_{S}=1$ while at $\beta=1$ we find $p_{L}=p_{S}\equiv p$.
In order for the mean degree to be constant, Eq.~(\ref{eq:expected_degree})
yields the distance-based probabilities \begin{subequations}\label{eq:connection_probabilities}

\begin{align}
p_{S}(\beta) & =\frac{1}{1+\beta(N-1-k)/k},\label{eq:conn_prob1}\\
p_{L}(\beta) & =\frac{\beta}{1+\beta(N-1-k)/k}=\beta p_{S}(\beta).\label{eq:conn_prob2}
\end{align}
\end{subequations}The short-range node degree $k_{S}$ follows a
binomial distribution $\mathcal{B}(k,p_{S})$ and the long-range node
degree $k_{L}$ follows a binomial distribution $\mathcal{B}(N-1-k,p_{L})$
where $\mathcal{B}(n,p)$ has probability mass function $f_{k}(n,p)={n \choose k}(1-p)^{n-k}p^{k}$.
A schematic explanation of the model is shown in Fig.\ \ref{fig:model_definition}.
A simple network generation algorithm is given as follows. Each node
$0\leq u\leq N-1$ connects to each of its $k/2$ rightmost short-range
neighbors with probability $p_{S}$. Afterwards, $m_{L}$ long-range
edges are drawn, where $m_{L}$ follows $\mathcal{B}(N(N-1-k)/2,p_{L})$.
For each long-range edge one chooses a source node $u$ uniform at
random from $[0,N-1]$. This node is then connected to a long-range
neighbor $v=(u+k/2+z)\mod N$ where the integer $z$ is drawn uniform
at random from the interval $[1,N-k-1]$. If an already existing edge
was chosen, repeat the procedure for this long-range edge. This algorithm
has complexity $\mathcal{O}(Nk+\expv{m_{L}})$ for sparse networks.
Open source implementations of the algorithm are available as C++/Python
packages \citep{maier_cmhrn_2018,maier_smallworld_2018}.

\begin{figure}
\includegraphics[width=0.9\columnwidth]{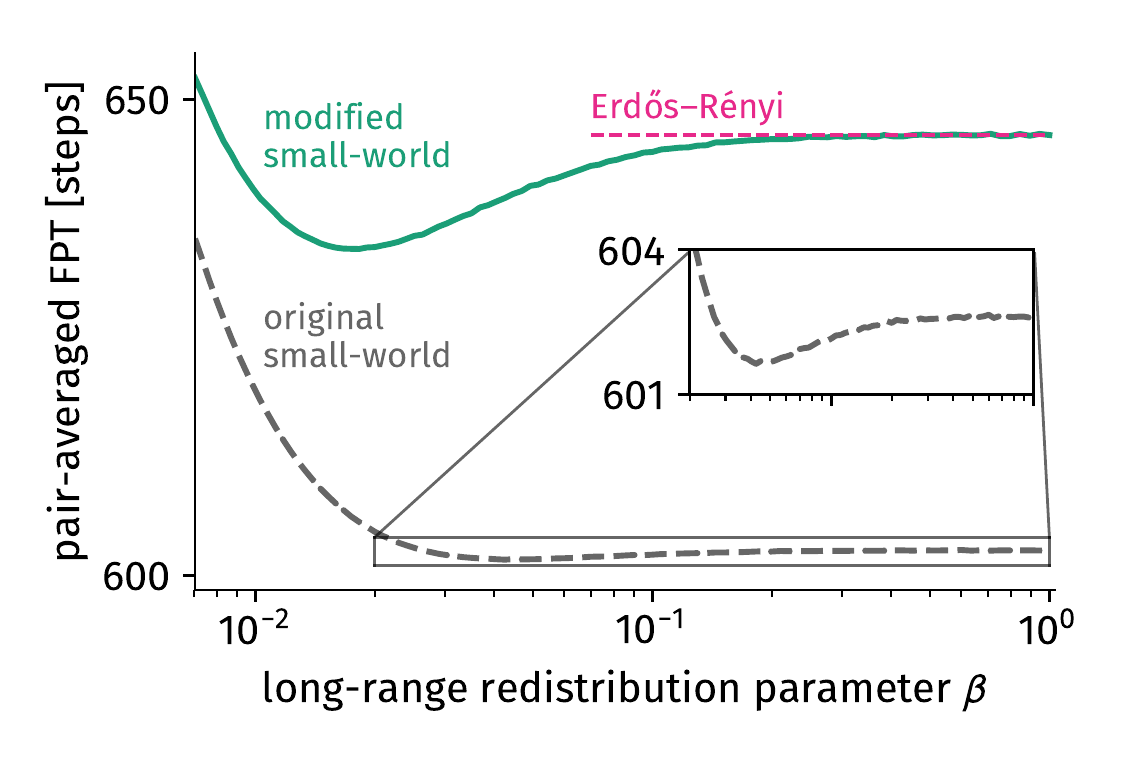}\caption{\label{fig:pafpt}The pair-averaged first passage time (PAFPT) of
a discrete-time random walk process is an example of a network observable
differing from the corresponding result of the \ER{} model in the
limit of $\beta=1$.In contrast, the result from the modified model
described in Sec.~\ref{sec:model_definition} approaches the desired
limit. Inset: A minimum in the PAFPT emerges in both the modified
as well as the original model, an effect explained in
\citep{maier_modular_2019}.}
\end{figure}

As the original model is widely used, we aim to highlight potential
consequences for the misinterpretation of the original model's randomized
limit in the following and compare it to the corresponding results
of the alternative model, which does approach the \ER{} model. To
this end it is first necessary to map the control parameters of the
two models in an appropriate manner such that they will be in similar
states when varying the parameters. We note that for small $\beta$
the short-range connection probability $p_{S}$ should be approximately
equal to the probability that an edge has \emph{not} been rewired
$1-p_{r}$ in the original model. To ensure that $p_{r}=1$ for $\beta=1$
we set $p_{r}=[1-p_{S}]\big/[1-p_{\mathrm{ER}}]$. In order to compare
the structural consequences on dynamic observables of both models,
we will compute a temporal observable of a discrete-time random walk
process as an example, a process defined as follows: At each discrete
time step a walker residing on a node $u$ chooses to jump to any
of $u$'s neighbors with uniform probability, repeated indefinitely.
Random walks are widely applied to model spreading and search processes
in physics, biology and computer science \citep{oksendal_stochastic_1992,berg_random_1993,barrat_dynamical_2008,newman_networks:_2010,klafter_first_2011,masuda_random_2016}.
Within this context, the mean first passage time $\tau_{vu}$ is the
expected number of steps a random walker needs to traverse to node
$v$ when it started at node $u$ which therefore can be interpreted
as an upper bound for any search process \citep{maier_modular_2019}.
In contrast, the shortest path length between two nodes is the search
time for an optimal search process. Based on the mean first passage
time, the pair-averaged first passage time $\tau=(N(N-1))^{-1}\sum_{v=1}^{N}\sum_{u\neq v}^{N}\tau_{uv}$
acts as a coarse-grained estimation of how fast a random search process
can be conducted between any two nodes of a particular network. We
computed the pair-averaged first passage time for small-world networks
of $N=512$ nodes and mean degree $k=10$. The control parameters
of both the alternative and the original model were varied ($\beta$
and $p_{r}=(1-p_{S}(\beta))/(1-p_{\mathrm{ER}})$, respectively).
For each value of $\beta$ we built the average of the pair-averaged
first passage time over the largest component of 10,000 independent
network realizations. For each realization, the mean first passage
times between all pairs of nodes of the largest component were computed
using Eq.~(14) in \citep{lin_mean_2012}. The results shown in Fig.~\ref{fig:pafpt}
imply that, indeed, the difference between both models can be of significance,
reaching values of a relative difference of up to $\approx7\%$ in
the randomized limit. This difference is induced by the fact that
in the original model, each node has a minimum degree of $k/2$ whereas
in the modified model nodes of smaller degree may exist \citep{maier_modular_2019}. 

The emergence of such a difference is an indicator for the relevance
of the alternative model for studying the influence of small-world
topologies on the outcome of dynamic systems -- the alternative model
is suited to compare its implications to the implications of a known
model, the \ER{} graph.

\subsection{Network properties of the alternative model}

\label{sec:model_properties}
\begin{figure}
\includegraphics[width=0.9\columnwidth]{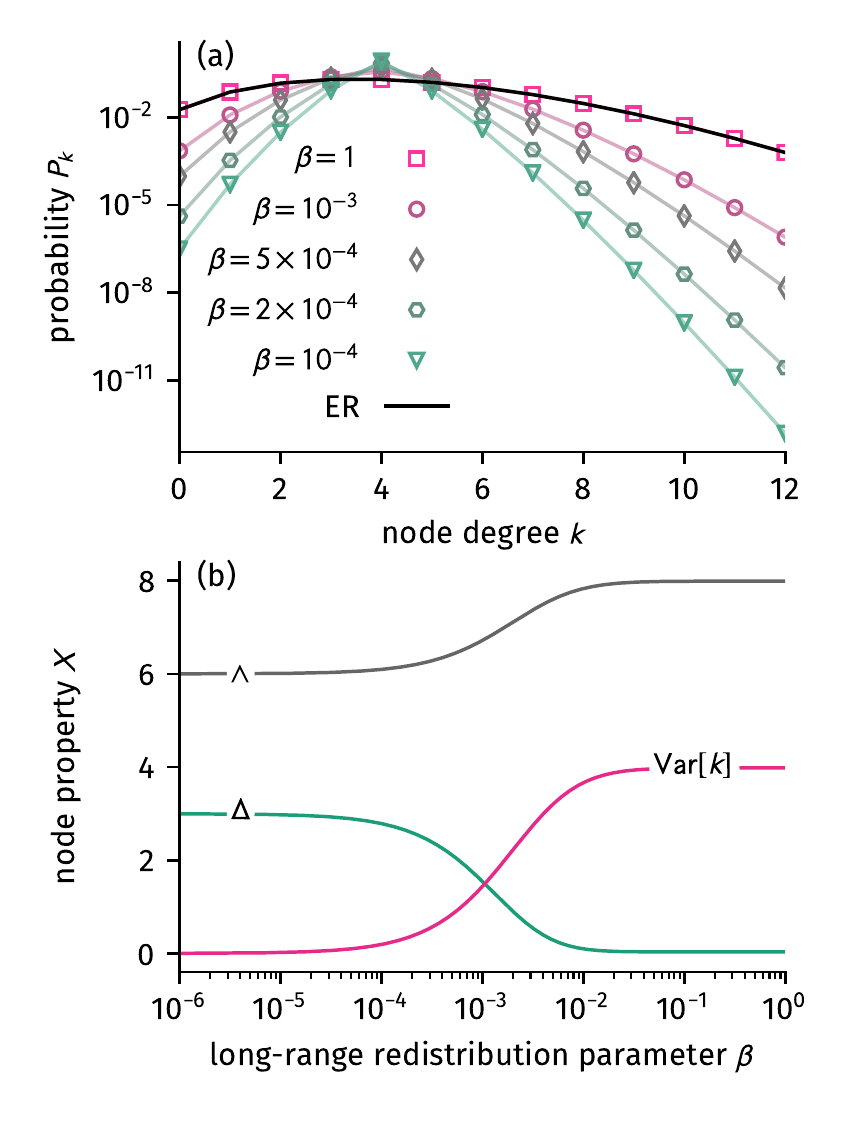}\caption{\label{fig:network_properties}Analytic results for (a) degree distribution
Eq.~\eqref{eq:degree_distribution} and (b), expected number of two-stars
per node $\wedge$, expected number of triangles per node $\triangle$
(Eq.~(\ref{eq:triangles})), and node degree variance as given in
the main text. While both degree variance and number of two-stars
increase with increasing long-range redistribution parameter $\beta$,
the number of triangles decreases. The results shown here were computed
for $N=1001$ and $k=4$.}
\end{figure}

We begin our discussion of the network
properties with the degree variance, which is important to quantify
the heterogeneity of nodes in a network based on their connectivity:
It has been shown that increased degree variance is increasing the
risk of endemicity of diseases on a network \citep{barrat_dynamical_2008}.
Furthermore, the degree variance plays an important role to estimate
the average arrival time of random walks \citep{maier_modular_2019}.
Because in the alternative small-world model the node degree is given
as the superposition of short-range and long-range degree, the degree
variance can be simply computed as 
\begin{align*}
\Var{k} & =\Var{k_{S}}+\Var{k_{L}}\\
 & =kp_{S}(1-p_{S})+(N-1-k)p_{L}(1-p_{L}).
\end{align*}
For increasing $\beta$ both short-range and long-range variances
increase, as well, such that the degree variance is an increasing
function of $\beta$, as shown in Fig~\ref{fig:network_properties}b.
The full degree distribution is computable by noting that any node
degree is $k_{i}=k_{S,i}+k_{L,i}$, such that its distribution is
given by the convolution
\begin{align}
p_{k'} & =\sum_{k_{S}=0}^{\infty}\sum_{k_{L}=0}^{\infty}f_{k_{S}}(k,p_{S})f_{k_{L}}(N-1-k,p_{L})\delta_{k',(k_{S}+k_{L})}\nonumber \\
 & =\sum_{k_{S}=0}^{\min(k',k)}{k \choose k_{S}}{N-1-k \choose k'-k_{S}}(1-p_{S})^{k-k_{S}}\times\nonumber \\
 & \qquad\qquad\times p_{S}^{k_{S}}(1-p_{L})^{N-1-k-k'+k_{S}}p_{L}^{k'-k_{S}},\label{eq:degree_distribution}
\end{align}
which is similar to the result derived in \citep{song_simple_2014}
and is shown in Fig.~\ref{fig:network_properties}a. Note that in
the derivation above we used Kronecker's delta $\delta_{ij}=0$ if
$i\neq j$ and $\delta_{ij}=1$ otherwise. Both the results of the
degree variance and the degree distribution highlight the simplicity
of the alternative model, which allows for a simple analytical evaluation
as compared to more complicated derivations in the original model
based on rewiring \citep{barrat_properties_2000}.

While there exist multiple similar definitions, the clustering coefficient
is usually reflecting the probability of triadic closure: Given a
structure where a node $i$ is connected to nodes $v$ and $u$, the
clustering coefficient is the probability that $u$ and $v$ are connected,
as well. Using the network's $(N\times N)$-sized adjacency matrix
$A_{ij}=1$ if nodes $i$ and $j$ are connected and $A_{ij}=0$ otherwise,
we therefore define the global clustering coefficient as the conditional
probability
\[
C=P\left[A_{iu}A_{uv}A_{vi}=1\Big|A_{iu}A_{iv}=1\right]=\frac{\expv{A_{iu}A_{uv}A_{vi}}}{\expv{A_{iu}A_{iv}}}\equiv\frac{\triangle}{\wedge},
\]
similar to the definition in \citep{song_simple_2014}. We will, however,
derive the final result using a more geometric approach in the following.
The probability $\wedge=\expv{A_{iu}A_{iv}}$ is the expected number
of two-stars per node (a structure where node $i$ is connected to
both a node $u$ and a node $v$). To evaluate this quantity one observes
that a node of degree $k_{v}$ is part of $(1/2)k_{v}(k_{v}-1)$ two-stars.Therefore,
it is given as $\wedge=(1/2)\left[\Var k+k(k-1)\right]$ . It hence
qualitatively follows the behavior of the degree variance as illustrated
in Fig.~\ref{fig:network_properties}b.
\begin{figure}
\raggedright{}\includegraphics[width=1\columnwidth]{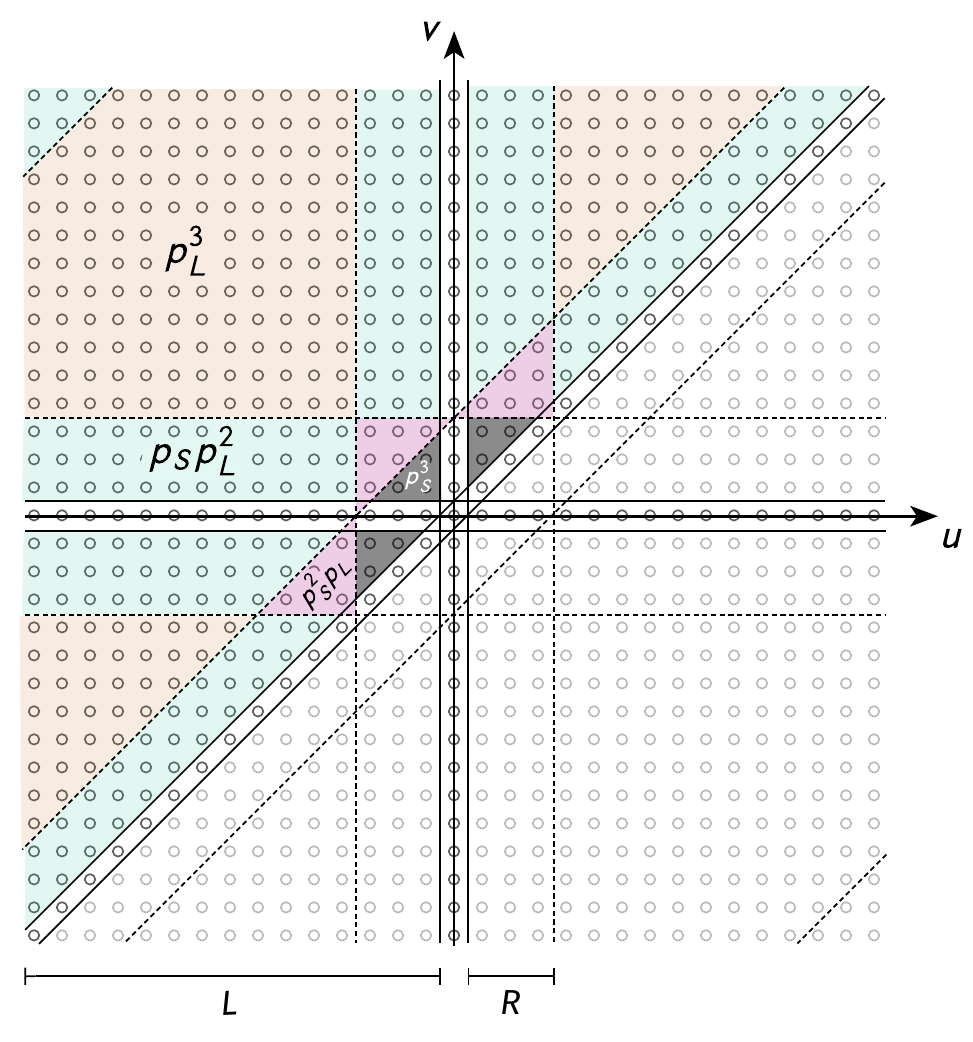}\caption{\label{fig:triangle-Evaluation}Evaluation of the areas of summation
to find the expected number of triangles per node $\triangle$ for
odd numbers of nodes $N$ as per Eq.~(\ref{eq:triangles}). Note
that here, the sum has been shifted to be $\sum_{u=-(N-1)/2+1}^{(N-1)/2}\sum_{v=u+1}^{(N-1)/2}(\cdot)$
such that $u$ and $v$ are equal to their lattice distance to a focal
node 0.}
\end{figure}

In order to find the expected number of triangles per node we recognize
that every node is statistically equivalent. Thus, without loss of
generality, we compute the number of triangles per node $i=1$ as
the sum over all possible remaining node pairs considering their
distance-based connection probability as 
\begin{align}
\triangle & =\sum_{u=2}^{N-1}\sum_{v=u+1}^{N}p_{d(u,1)}p_{d(v,1)}p_{d(u,v)}\nonumber \\
 & =Fp_{S}^{3}+Gp_{S}^{2}p_{L}+Hp_{S}p_{L}^{2}+Ip_{L}^{3}.\label{eq:triangles}
\end{align}
Here, $F,$ $G$, $H$, and $I$ are the areas of summation highlighted
in Fig.~\ref{fig:triangle-Evaluation} where three (grey), two (pink),
one (green), and no (orange) node pairs are of short-range distance,
respectively . Considering the case of odd numbers of $N$ one may
shift the summations to run from lattice distance $-N/2$ to distance
$N/2$ around a focal node at $d=0$ such that by marking the conditions
for short-range connections, finding the respective areas reduces
to a geometrical exercise. By defining the lengths $L=(N-1)/2$ and
$R=k/2$ as marked in Fig.~\ref{fig:triangle-Evaluation}, one first
finds the useful unit of a short-short-long-range area as the triangle
$T=(R^{2}-R)/2+R$ (marked as pink in Fig.~\ref{fig:triangle-Evaluation}).
Then, the areas of summation are given as
\begin{align*}
F & =3(T-R)=(3k/8)(k-2)\\
G & =3T=(3k/8)(k+2)\\
H & =2((L-R)R-T)+T+2(L-R)R+2((L-1)R-T)\\
 & =(k/8)(12N-26-11k)\\
I & =(L-R)^{2}-2((L-1)R-T)-(L-R)+(L-R)^{2}-T\\
 & =(1/8)\left[5k^{2}-k(12N-26)+4(N^{2}-3N+2)\right].
\end{align*}
The expected number of triangles Eq.~(\ref{eq:triangles}) consequently
decreases with increasing $\beta$, as expected and as shown in Fig.~\ref{fig:network_properties}b.
Considering Eqs.~\eqref{eq:connection_probabilities}, the exact
value of the clustering coefficient is then given by
\begin{equation}
C(\beta)=p_{S}^{3}\times\frac{F+G\beta+H\beta^{2}+I\beta^{3}}{(1/2)\Var k+k(k-1)}.\label{eq:clustering_exact}
\end{equation}
In the respective limits we find
\begin{align*}
C(\beta=0) & =\frac{3(k-2)}{4(k-1)}\\
C(\beta=1) & =\frac{\sum_{u=2}^{N-1}\sum_{v=u+1}^{N}p^{3}}{\sum_{u=2}^{N-1}\sum_{v=u+1}^{N}p^{2}}=p,
\end{align*}
which are the expected results for both the $k$-nearest neighbor
lattice as well as the \ER{} graph. Further considering Eqs.~\eqref{eq:connection_probabilities}
and \eqref{eq:clustering_exact} as well as noting that $\Var{k}(\beta\rightarrow0)=0$,
in the limit of small long-range redistribution one finds
\begin{align}
\frac{C(\beta\ll1)}{C(0)} & \approx p_{S}^{3}=1-3\beta\frac{N-k-1}{k}+\mathcal{O}(\beta^{2})\label{eq:clustering-scaling}
\end{align}
which will be of importance for quantifiying the small-world effect
in the following.

\subsection{Small-world effect}

\label{sec:small-world-effect}In the original model, the small-world
effect was illustrated by comparing the clustering coefficient to
the average shortest path length of networks. While random networks
have short path lengths, they possess low clustering, on the other
hand regular networks are highly clustered, while nodes are, on average,
quite distant from one another. With rewiring only a short amount
of edges of an ordered network it was shown that shorter paths appear
immediately while high clustering preserves, explaining the small-world
effect. It has further been argued that algorithmic searches requiring
local information are necessary to identify these short paths \citep{kleinberg_small-world_2000,watts_identity_2002}.
However, in situations where searches are less targeted and follow
rather diffusive dynamics such as epidemic spreading over air traffic
\citep{iannelli_effective_2017} or synchronization in oscillators
\citep{barrat_dynamical_2008}, the role of the mean shortest path
length becomes less prominent. Rather, random walk relaxation and
passage times are the important observables characterizing these dynamics,
specifically to predict the arrival time of a disease or the likelihood
of global synchronization. Therefore, we will take an approach focusing
on random walks in the following.

One of the purposes of the original model was to explain the Milgram
small-world experiment \citep{travers_experimental_1969} where participants
had to mail letters to strangers by mailing them to a person they
did know and instruct them to pass the letter further. In the following
we will illustrate the small-world effect by showing that an upper
bound for the delivery time of those messages decreases much faster
than the clustering coefficient with increasing probability of long-range
edges. Since this upper bound of a random search also bounds the mean
shortest-path length which is the equivalent to the arrival time of
a maximally informed search, the following result generalizes the
small-world effect to random dynamics.

\begin{figure}
\includegraphics[width=1\columnwidth]{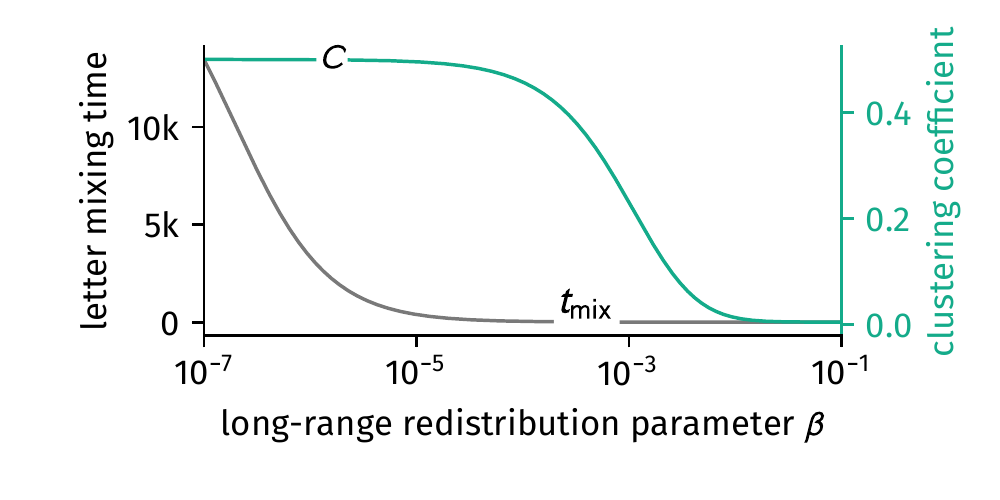}\caption{\label{fig:small-world}The small-world effect as illustrated by the
observables computed analytically in this paper. The random walk message
delivering mixing time Eq.\ \eqref{eq:mixing_time_exact} as an upper
bound of targeted search message delivering mixing time decreases
rapidly with the introduction of long-range links while clustering
Eq.\ \eqref{eq:clustering_exact} preserves. Displayed here are results
for $N=1001$ and $k=4$.}
\end{figure}

Considering completely uninformed individuals, the mailing process
is modeled as a random walk process where the random walkers correspond
to the letters to be sent to recipients. At each integer time step
$t$, the letter resides on a node $u$. Subsequently, one of $u$'s
neighbors $v$ is chosen uniform at random as the next recipient of
the message. At the next time step $t+1$ the letter then resides
at node $v$. This process is repeated indefinitely and is governed
by the master equation $\varphi_{v}(t)=\sum_{u=1}^{N}(A_{vu}/k_{u})\varphi_{u}(t-1)$
where $\varphi_{v}(t)$ is the probability that the letter is on node
$v$ at time $t$ and $W_{vu}=A_{vu}/k_{u}$ is the probability that
the letter is sent from node $u$ to node $v$. Instead of generating
adjacency matrices and averaging over the results of their corresponding
transition matrices we will compute an average medium matrix where
each edge in the network is replaced by the probability of this edge
existing such that $W_{vu}^{\mathrm{avg}}=\expv{A_{vu}}/k=p_{d(v,u)}/k$.
One can show that the time scale with which the equilibrium distribution
is approached on this average medium network is given by the eigenvalue
gap of the transition matrix $W_{vu}^{\mathrm{avg}}$ as $t_{\mathrm{mix}}^{-1}=1-\omega_{1}$
where $\omega_{0}=1$ is the largest eigenvalue and $\omega_{1}$
is the second largest eigenvalue \citep{hahn_applications_1997}.
The average medium transition matrix $W_{vu}^{\mathrm{avg}}$ is circulant
based on the vector 
\[
w=k^{-1}(0,\underbrace{p_{S},\dots,p_{S}}_{k/2},\underbrace{p_{L},\dots,p_{L}}_{N-1-k},\underbrace{p_{S},\dots,p_{S}}_{k/2}).
\]
In this case, the $j$-th eigenvalue of $W_{vu}^{\mathrm{avg}}$ is
given as $\omega_{j}=\sum_{v=1}^{N}w_{v}\exp(i2\pi v/N)$ such that
the second largest eigenvalue can be easily computed as $\omega_{1}=p_{S}\Gamma/k-p_{L}(1+\Gamma)/k$
where $\Gamma=2\sum_{j=1}^{k/2}\cos(2\pi j/N)=k-(\pi/N)^{2}k(k/2+1)(k+1)/3+\mathcal{O}(N^{-4})$
which yields the mixing time
\begin{equation}
t_{\mathrm{mix}}(\beta)=\left[1-\frac{\Gamma-\beta(1+\Gamma)}{k+\beta(N-1-k)}\right]^{-1}.\label{eq:mixing_time_exact}
\end{equation}
In Fig.\ \ref{fig:small-world} we show how both clustering coefficient
and mixing time decrease with increasing long-range redistribution
parameter $\beta$. In the limits we find the expected results from
$k$-regular networks and an average medium approximation of the \ER{}
graph 
\begin{align*}
t_{\mathrm{mix}}(\beta=0) & =\left[1-\frac{\Gamma}{k}\right]^{-1}\stackrel{N\gg k/2}{\longrightarrow}\frac{N^{2}}{\pi{}^{2}}\frac{3}{(k/2+1)(k+1)},\\
t_{\mathrm{mix}}(\beta=1) & =\left[1-\frac{1}{N-1}\right]^{-1}=1-\frac{1}{N}.
\end{align*}
This implies that for small long-range redistributions the relative
mixing time decreases as 
\begin{equation}
\frac{t_{\mathrm{mix}}(\beta)}{t_{\mathrm{mix}}(0)}=1-\beta\left(\frac{3N^{3}}{\pi^{2}\left(k/2+1\right)(k+1)k}-\frac{N}{k}+\frac{1}{k}\right)+\mathcal{O}(\beta^{2}).\label{eq:mixing-time-scaling}
\end{equation}
Comparing Eqs.\ \eqref{eq:clustering-scaling} and \eqref{eq:mixing-time-scaling},
one can easily see that for small $\beta$ the rate with which the
mixing time decreases is of order $N^{3}$ while the rate with which
the clustering coefficient decreases is of order $N$, which is a
difference of two orders of magnitude. This shows that even with a
small amount of long-range connection probability, the delivery time
of randomly passed messages declines rapidly while clustering is still
preserved. Since an optimal search strategy identifies the shortest
path between two nodes and the original small-world effect was shown
for those shortest paths, this result generalizes the small-world
effect to random search strategies.

\section{Discussion}

\label{sec:conclusions}We discussed an alternative small-world network
model first introduced in Ref.~\citep{song_simple_2014}, which approaches
the \ER{} random graph model in the limit of maximum disorder and
showed that the original small-world network model does not. Within
this model, instead of rewiring edges, long-range contacts are introduced
by redistributing connection probability from short-range to long-range
potential neighbors while keeping the mean degree constant. Constructing
small-world networks in this way allows for a thorough analytical
analysis of network properties such as the degree distribution, the
degree variance, the average number of two-stars, the average number
of triangles, and the clustering coefficient. An upper bound of the
message deliviery time can be computed using an average medium approximation.
We showed that for a small amount of redistributed long-range connection
probability the clustering coefficient decreases with a rate proportional
to the number of nodes $N$ while the upper bound of the delivery
time decreases with a rate of order $N^{3}$, hence illustrating how
social networks can have both high clustering as well as a favorable
topology to efficiently forward messages to unknown recipients, even
if the search strategies are purely random, as they might be in diffusive
contexts such as epidemic spreading in air traffic or synchronization
of oscillators.

In the following we will discuss the modified model's applicability
to teach network concepts. As network theory curricula typically introduce
\ER{} random graphs early on as one of the first network models,
the concept of drawing edges with a certain probability is known to
students. We argue that extending this concept to draw edges from
two categories (short-range and long-range) with two connection probabilities
is a natural way to extend this formalism on a path to more complicated
models. Based on the derivation of the degree distribution of the
random graph one can easily comment on the distribution of random
variables' superposition and derive the degree distribution of the
small-world model. Subsequently, similarly to the clustering coefficient
computable in the \ER{} model, the clustering coefficient of the
modified small-world model can be computed as the conditional probability
that two nodes are connected given that they are neighbors of a focal
node, in contrast to the local clustering coefficient in the original
model. This further allows for the introduction of an average medium
where each edge is replaced by the probability that it exists. Consequently
using this average medium approximation one can use the modified model
to introduce the random walk formalism and show how to evaluate its
mixing time to arrive at the small-world effect based on the message
delivery time and the Milgram small-world experiment (with a careful
discussion of its flaws). We furthermore argue that the more simplistic
picture of drawn instead of rewired edges is more intuitive. Instead
of an individual explicitly deciding to change one of its short-range
contacts to a long-range contact, there is an inherent probability
to be connected to ``near'' nodes as well as a smaller probability
to be connected to nodes ``further away''.

Finally, we suggest the modified model to be used as an alternative
to the original model when studying the influence of the small-world
effect on dynamic systems, since the modified model truly interpolates
between two well-studied systems, a nearest-neighbor lattice and the
\ER{} model. It therefore allows for simpler and more reliable comparisons
of results and potentially offers more insight to other dynamics due
to its analytical accessibility.

\section*{Materials}

The network sampling algorithm described in
Sec.~\ref{sec:model_definition}
is implemented for Python and C++ and available for download \citep{maier_cmhrn_2018,maier_smallworld_2018}.
Additionally, several Python functions to compute the model's network
properties as well as the average medium mixing time are implemented
in \citep{maier_smallworld_2018}.

\section*{Data availability}

No datasets were generated or analysed during the current study.

\section*{Competing interests }

The author declares no competing interests.

\section*{Acknowledgments}

B. F. M. wants to thank D. Brockmann, B. Sonnenschein, M. Wiedermann,
F. Ianelli, and A. Schwarze for helpful comments. The author is financially
supported as an \emph{Add-On Fellow for Interdisciplinary Life Science}
by the Joachim Herz Stiftung.

\end{document}